# Resistive switching effect in the n-InGaAs/GaAs heterostructures with double tunnel-coupled quantum wells

P.A. Belevskii, M.N. Vinoslavskii, V.V. Vainberg, O.S. Pylypchuk, V.N. Poroshin

Institute of Physics, National academy of sciences of Ukraine

46 Prospekt Nauki, Kyiv, 03028 Ukraine

E-mail: pylypchuk@iop.kiev.ua

**Abstract**

The electric conductivity behavior in the single and double tunnel-coupled quantum wells (QW) with different doping profile caused by impact of short pulses of the strong longitudinal (in the quantum wells plane) electric field has been investigated. It is established that at low temperatures (4 K) after such impact the long-term metastable state with increased electric conductance may be observed in the case of the asymmetric QW couple with the impurity delta shaped layer in the narrower QW. It is not observed in the structures with other configurations. The observed effect is explained by the model accounting for metastable changes in the electron energy states spectrum in the studied structures caused by the strong electric field pulses.

**Introduction**

Resistive switching, i.e., change of the electric resistivity due to impact of short electric field pulses was observed in different materials and structures, such as metal-dielectric-metal and metal-dielectric-semiconductor with different dielectrics. In particular, it was observed in the binary transition metal oxides (see, for example, [1]), chalcogenides [2], organic materials [3], amorphous Si (a-Si) [4] and graphene oxide [5]. Such resistive switching is usually reversible.

To explain causes of the resistive switching observed in these structures several physical phenomena are considered which take place in the dielectric bulk or at its interface with a conductor. These phenomena are formation and destruction of conducting channels [6], influence of the defect electromigration on the contact Schottky barrier resistance [7], injection and emission of charge carriers [8], band potential profile changing due to capture of charge carriers by traps in the dielectric bulk and surface [9,10], etc. The results of investigation of the resistive switching and its mechanisms in different structures are reviewed in [11].

The present paper reports on finding and investigation of the resistive switching effect in the multi-layer semiconductor InGaAs/GaAs heterostructures with asymmetrically doped double and tunnel-coupled quantum wells (DTCQW) in the lateral conduction regime.

**Studied structures and experimental details**

The studied heterostructures of $In_xGaAs_{1-x}$/GaAs, x= 10 %, were grown by the MOVPE method on the semi-insulating GaAs substrates[1]. In one group of structures each one consists of 10 single quantum wells (QW) being 80 Å wide and separated by the 800 Å wide barriers. They were doped by the delta-shaped layer in the QW center or in the barrier at 100 Å apart of QW. The impurity concentration was $\sim 2 \cdot 10^{11}$ cm$^{-2}$ per period. The background impurity concentration was $2 \cdot 10^{15}$ cm$^{-3}$. Another group consisted of structures containing 50 double QW (DTCQW) 70 and 170 Å wide with the 50 Å wide GaAs barrier between them. The barrier between couples was 800 Å wide. The wide wells were undoped while the narrower ones were delta-doped by silicon with concentration of 1.1 through $3.4 \cdot 10^{11}$ cm$^{-2}$ per period.

The samples for measurements were supplied by the Ohmic contacts made by deposition and alloying Ge/Au. Two kinds of samples were used. The first kind samples had 2 electric contacts with the distance of about 0.3 cm between them. They were used for measurement of electric conduction only. The second kind samples were shaped as the Hall bridges and used to measure both the electric conduction and electron concentration simultaneously. The initial electric conductivity and electron concentration were measured in the dc regime at 4.2 K at the electric field strength of 10 V/cm. Afterward, the single voltage pulses in the range of 20 through 1000 V with duration of 0.5 μs were applied to the current contacts. The waveforms of the voltage and current through the sample were registered by the digital oscilloscope Tektronics TDS 1002 connected to the computer. After measurement in the pulse regime the small dc voltage was applied again to the sample and the electric conductivity and electron concentration by means of the Hall effect were measured several times after different delays.

---

[1] The samples were fabricated in the Physical-Technical Research Institute, N.I. Lobachevsky State University, Nizhny Novgorod, Russia.

All measurements were carried out at 4.2 K. Cooling of samples from the room temperature down to 4.2 K was performed keeping samples without lighting.

**Measurement results**

The electric conductivity vs electric field dependence manifests qualitatively different behavior in the studied kinds of structures (figure 1). In the structures delta-doped in the barrier the electric conductivity decreases with the electric field. On the contrary, in the structures with single QW delta doped in its center the conductivity increases with growing field.

The conductivity of the structures with asymmetrically δ-doped DTCQW has a more complicated character. First, with the electric field growing it increased and afterward with further field increase it somewhat decreased.

In all structures, except of the asymmetrically doped DTCQW, the conductivity after turning off the pulsed electric field returned to its initial value $\sigma_0$ measured before the electric field impact. In the case of DTCQW structures the conductivity decrease occurred down to some steady-state value exceeding the value of $\sigma_0$ (figure 2). The excess conductivity value depends on the electric field magnitude and structure doping level. And it remained unchanged during long period at low temperature.

Shown in figure 3 and figure 4 are the conductance vs electric field and Hall electron concentration vs electric field dependences, respectively, measured in the dc regime after turning off the pulsed electric field.

The obtained results indicate the strong pulsed electric field in these structures to cause switching the low conductivity state into the high conductivity one existing long term, no less of several hours, after turning off the voltage pulse. Notice, that the switching effect is irreversible with the field. The conductivity magnitude returns to its initial value of $\sigma_0$, if the sample is heated at least up to 100 K and then is cooled again down to the measurement temperature.

**Discussion**

The observed peculiarities of the conductivity behavior in the structures with different location of the impurity layer may be explained by difference in spectra of the electron states participating in conduction and their filling by electrons. In the case of a single well structure with the δ-impurity layer in the barrier this layer forms an additional quantum well in the barrier with the size quantization levels higher by energy for the used doping level than in the main QW [12]. At the liquid helium temperature all electrons fill the states of the main well. With increasing electric field, the electrons are heated up to the level in the δ-impurity layer well and transferred to it by means of tunneling [13]. The total mobility of electrons decreases due to scattering by ionized impurities and, consequently, the conductivity decreases. After turning off the field pulse the electrons return into the main well by tunneling and conductivity returns to the initial one $\sigma_0$.

In the case of the structures with the single QW doped in the well center the low temperature conduction is caused mainly by electrons in the two-dimensional impurity band located by energy below the first size quantization subband [14]. The electric field transfers electrons from the impurity band, where their mobility is small, into the two-dimensional QW subband with the much higher electron mobility. Consequently, the conductivity increases in accordance with experiment. After turning off the pulsed field the excited electrons quickly relax by energy and return into the impurity band, and the conductivity recovers to the value of $\sigma = \sigma_0$.

Let us now consider the energy spectrum of electrons in the structures with DTCQW. Since the δ-layer of impurity is in the narrower well, the impurity band is bound to the size quantization subband of this well. We carried out simulation of the energy spectrum of electrons in this case, including the spatial distribution of the electron density, defined by the squared envelop wavefunctions, by means of self-consisted solution of the Schrodinger and Poisson equations along the structure growth axis with the electroneutrality condition per period. The spatial distribution of the impurity atoms was set in the gaussian form with the mean squared deviation σ = 10÷20 Å evaluated from the known technological parameters. The distribution of the impurity states density by energy was also fitted by the gaussian with $\sigma_N \approx c\frac{e^2}{\kappa}\sqrt{N}$, where c is the coefficient by order of unity, e is the electron charge, κ is the dielectric permittivity. The impurity band width estimate for our samples, defined by inter-impurity distance, accounting also for its fluctuation and composition fluctuation, give 4 - 20 meV (see, for instance, [15]). The typical view of this kind simulated energy spectrum in the equilibrium state at 4.2 K is depicted in figure 5 **a**.

Due to difference in the energy spectra in the narrow and wide wells the impurity band center in the narrow band is close by energy to the bottom of the size quantization subband in the wide well. The electron states in both bands partly overlap by energy. Due to this fact and high density of states in the impurity band a quite large portion of electrons is at the helium temperature in the impurity band. The larger impurity concentration, the larger portion of electrons in the impurity band. In the high electric field, the electron energy in both wells increases leading to their mobility increase because the ionized impurity scattering dominates at the temperature of experiment (in the undoped well because of the long distance action of the Coulomb potential of the impurity in the narrower well). Along with it the electrons may transfer from the impurity band states with a low mobility into the states of the first subband in the wide well with the higher electron mobility. This transition may occur via two channels. One occurs by tunneling from the first subband of the narrower well where the electrons transfer to from the impurity band due to heating by the electric field. The second channel is the direct tunneling from the impurity band states into the first subband states in the wide well which become empty in the course of heating electrons. Such transition is possible because of overlapping spectra of electron energy in coupled wells. In consequence the structure conductivity increases.

After the field is turned off the conduction in such structures, as mentioned above, relaxes down to some value exceeding the equilibrium thermodynamic magnitude. One may suppose the following reason of this phenomenon. The transition of electrons during impact of the electric field from the impurity band into the wide well leads to change of the spatial charges in coupled wells and internal electric field between them which exists in the asymmetrically δ-doped structures. This, in turn, changes the potential profile by shifting it in the opposite directions in the wells and respectively shifting electron energy states in both wells. As a result, overlapping of energy states in the impurity band and those in the first subband in the wide well may decrease or disappear at all. Therefore, the backward transition of electrons from the wide well into the narrow one becomes impossible and the electron concentration in the wide well with the higher mobility after the pulse is turned off remains larger than initial equilibrium one in this well. As a result, the conductivity exceeds the initial value. Note, the electron concentration increase in the wide well in result of the electric pulse impact is confirmed by results of the Hall effect measurements (figure 4).

To confirm the proposed model, we calculated the metastable potential profile and electron energy spectrum for the structure in figure 5**a**, supposing the 25 % of electrons transferred from the impurity band into the first subband in the wide well due to the electric field impact. The calculation results are shown in figure 5**b**. Comparing figure 5 **a** and **b**, one sees that transition of electrons indeed leads to moving downward the size quantization subband and impurity band in the narrow well and, respectively, the impurity band becomes lower the first subband in the wide well. As a result, overlapping of their energy states disappears and transition of electrons between them becomes impossible.

Note, the return of electrons into the impurity band may go via the size quantization subband of the narrow well. It demands for the electron energy in the wide well to be no less the energy of this narrow well subband which may be achieved, for instance, by heating sample up to temperatures higher than ~100 K.

**Conclusion**

Thus, we observed the effect of switching from the small conductivity state into the high conductivity state in the multi-layer InGaAs/GaAs heterostructures with double asymmetrically doped tunnel-coupled quantum wells at 4.2 K due to impact of the short longitudinal electric field pulses. This switching is caused by transition of electrons from the states of the impurity band in the narrow doped QW with small electron mobility into the sates of the first size quantization subband in the wide undoped well with the higher electron mobility. Impossibility of the backward transition is explained by change in the electron energy spectra in both wells in the course of switching which leads to absence of energy overlapping of the electron states in neighbor wells and makes impossible tunneling between them. This effect is not observed in other studied structures, such as those with the δ-layer of impurity in the barrier or in the center of a single quantum well.

We note, the proposed mechanism of change in the of electron states characteristics in DTCQW is similar to that considered in [16] for the DX centers in the δ-doped GaAs where it was explained by recharging DX states in the strong electric field leading to transition of the DX levels from the range of conducting states into the range of localized states.

**Acknowledgements**

The authors thank Dr N. Baidus for the submitted heterostructures and Prof. O. Sarbey for discussion on the results.

The work was supported by the target program for fundamental research of the Nat. Acad. Sciences of Ukraine "Perspective fundamental research and innovative development of nanomaterials and nanotechnology for needs of industry, health and agriculture", project 10/21-H.

Figure 1.

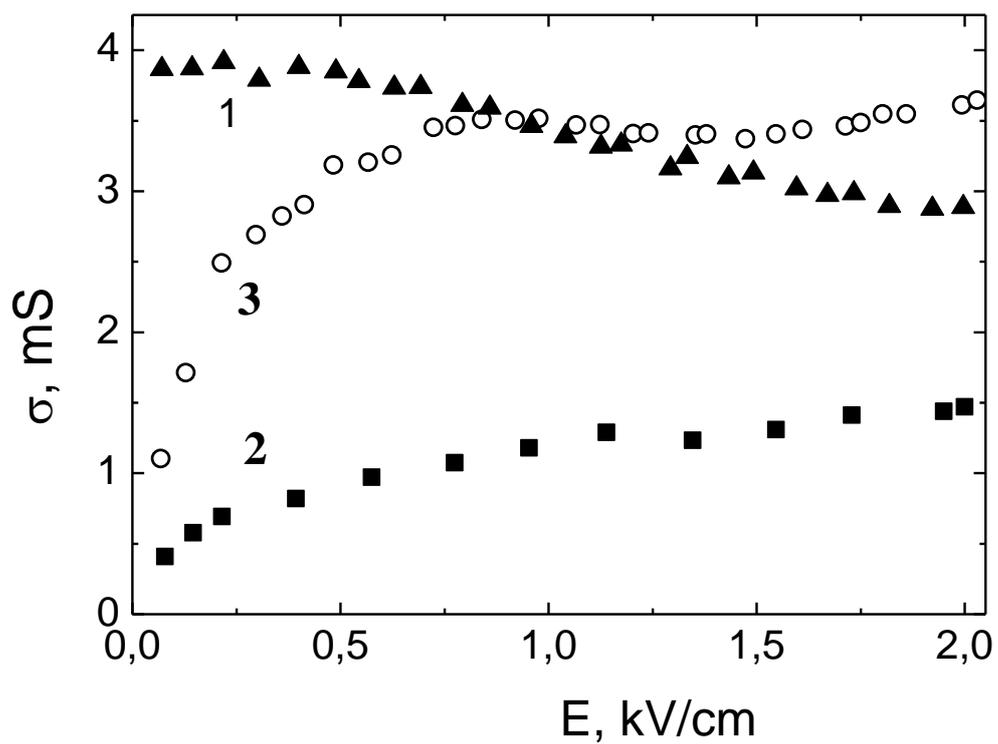

Figure 2.

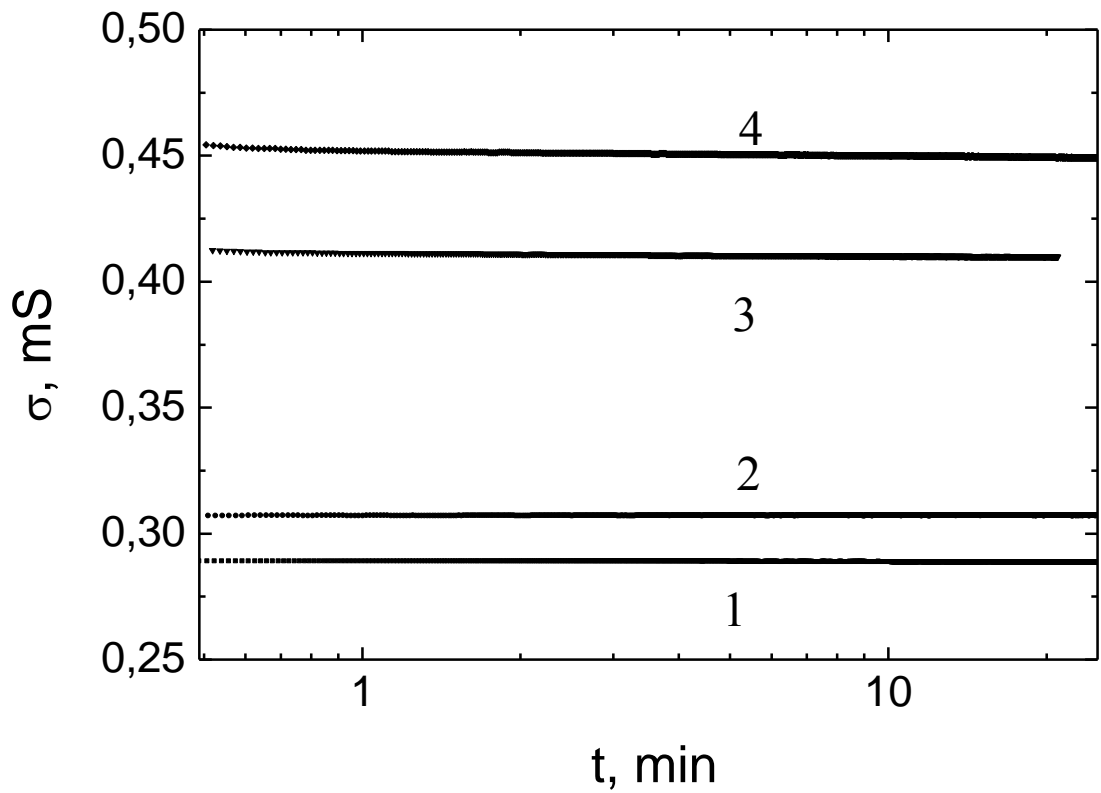

Figure 3.

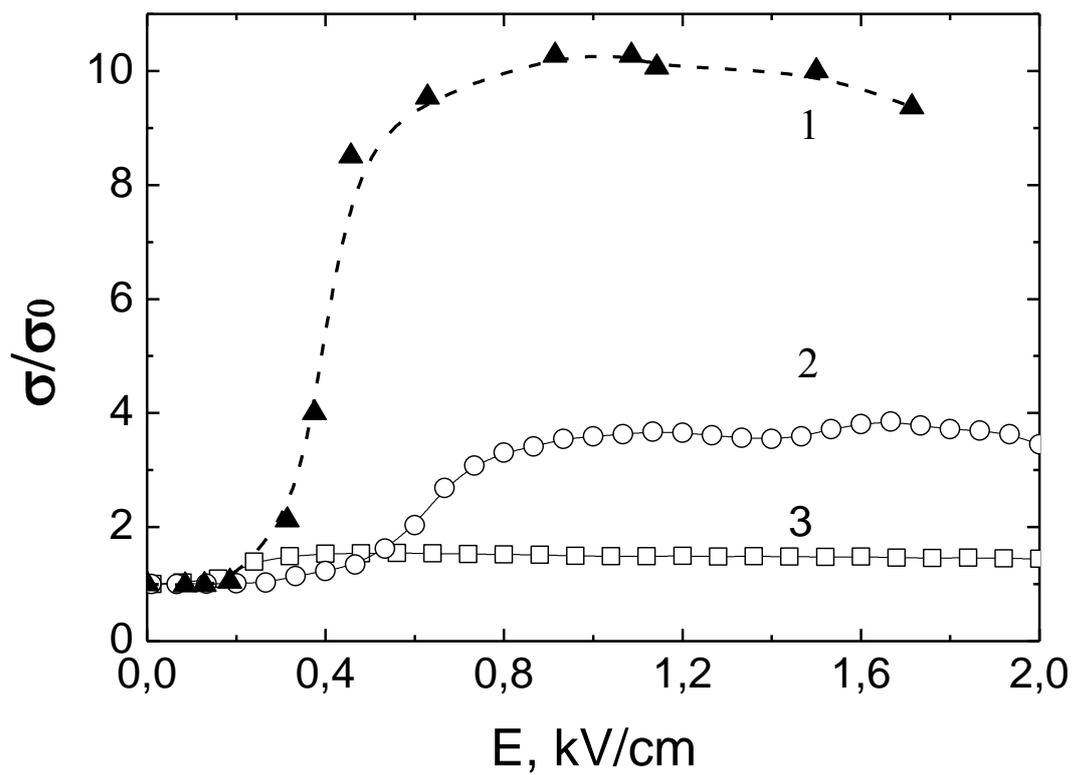

Figure 4.

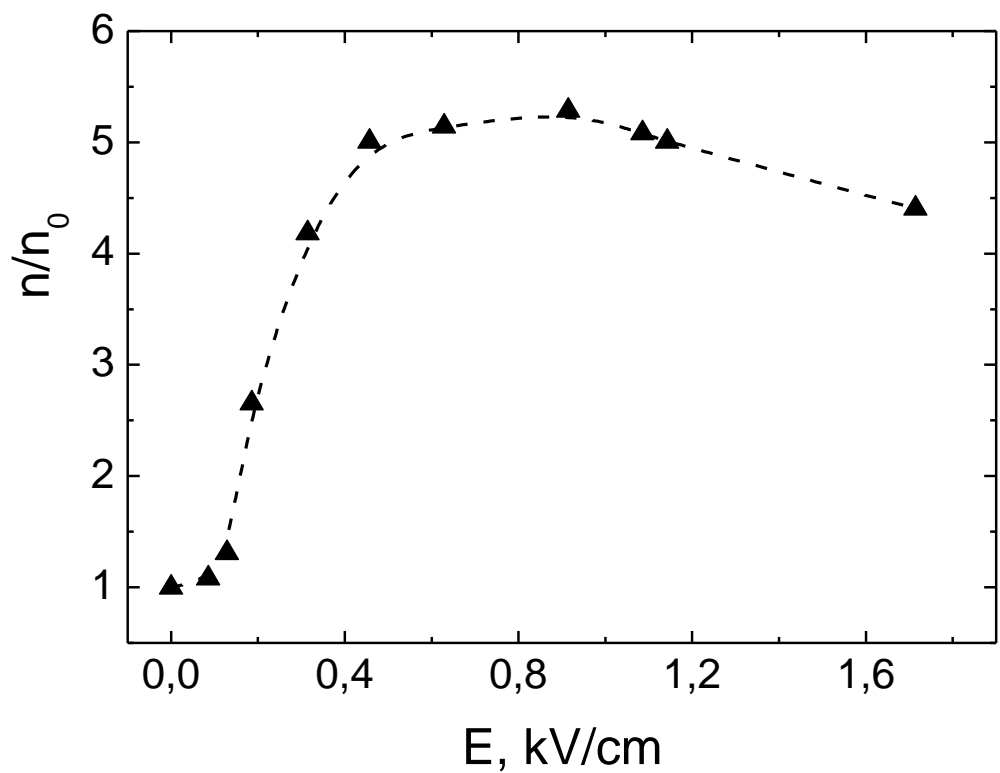

Figure 5.

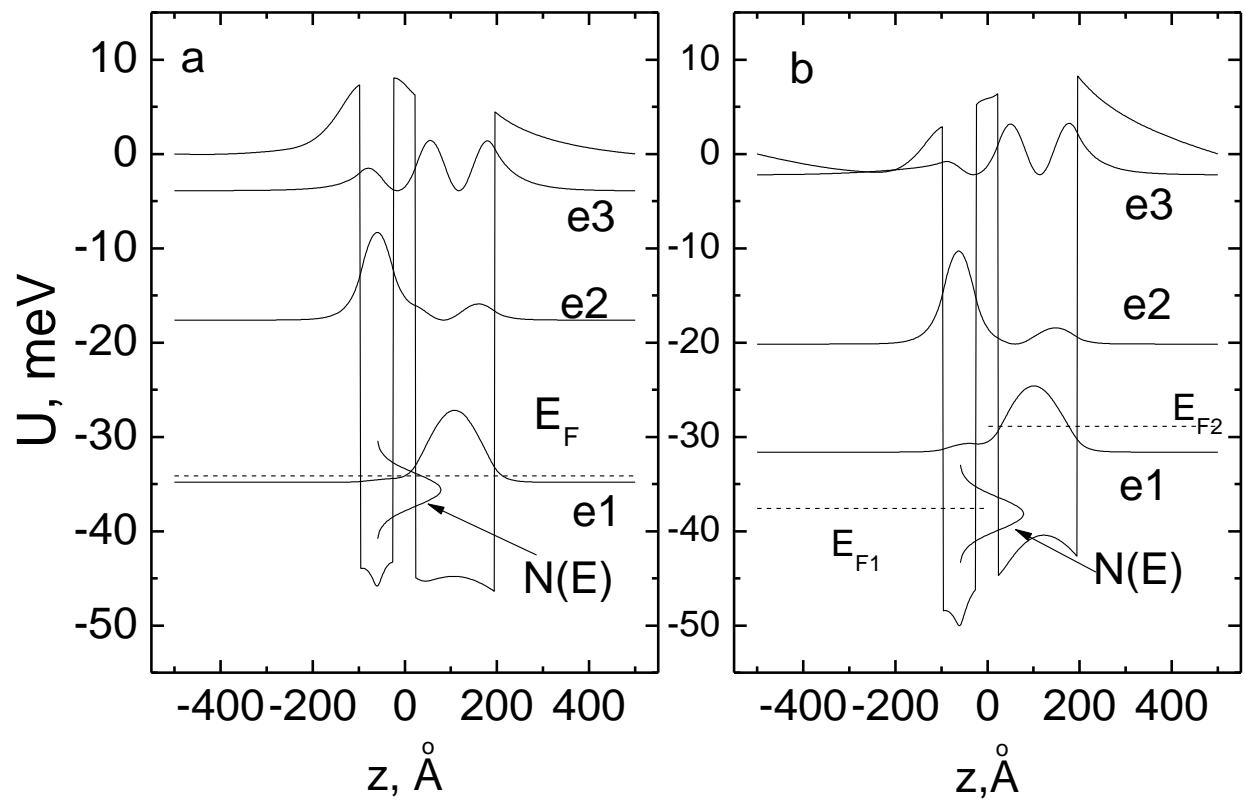

**Figure captions**

Figure 1. The electric conduction vs the pulsed electric field strength magnitude for the structures: 1 - with delta-doped barriers, 2 – delta-doped single QW, 3 – asymmetrically doped DTCQW. T=4.2 K. The impurity concentration, N, $10^{11}$ cm$^{-2}$: 1 -3.4, 2 and 3 – 2.

Figure 2. Conduction behavior with time in the DTCQW structures after turning off the electric field pulse of different magnitude F, kV/cm: 1 – 0.01 (initial $\sigma_0$), 2 – 0.17, 3 – 2.32, 4 – 2.8. The impurity concentration, N =·1.1·$10^{11}$ cm$^{-2}$.

Figure 3. The relative conductance vs pulsed electric field magnitude dependence in the DTCQW structures measured after turning off the pulsed electric field. The impurity concentration, N, $10^{11}$ cm$^{-2}$: 1- 3.4; 2-1.9; 3- 1.1.

Fgure 4. The Hall electron concentration vs pulsed electric field magnitude dependence in the DTCQW structures measured after turning off the pulsed electric field. The impurity concentration, N = 3.4·$10^{11}$ cm$^{-2}$.

Figure 5. The energy spectra of the structures with asymmetrically doped DTCQW (numerical calculation). The conduction band bottom profile and the squared electron envelop wave functions profiles: (**a**) in the equilibrium state before and (**b**) in the metastable state after impact by the strong longitudinal electric field pulses. e1, e2,e3 - the size quantization levels, N(E)- the density of states in the impurity band, $E_F$- the Fermi level in the equilibrium state, $E_{F1}$, $E_{F2}$- the quasi Fermi levels in the corresponding wells in the metastable state. The delta-impurity concentration N=2.57·$10^{11}$ cm$^{-2}$, T=4.2 K.